\documentclass[twocolumn,superscriptaddress,prl, showpacs]{revtex4}
\usepackage{amsmath,amssymb}
\usepackage{bm}

\begin{document}
\title{Parity Violation in Chiral Symmetry Breaking and Neutrino}

\author{Hidenori TAKAHASHI} \email{htaka@phys.ge.cst.nihon-u.ac.jp}
\affiliation{
 Faculty of Science and Technology, Nihon University,
 Japan 
}%

\date{\today}

\begin{abstract}
\ \\
I consider parity violation of the chiral symmetry broken vacuum.
It is shown that the parity of the true vacuum in the chiral symmetry breaking model is violated.
I propose new mechanism of massive neutrino, which
the neutrinos have mass in terms of the chiral symmetry breaking.
In this mechanism,  it may give us new understanding why almost neutrinos have left chirality.
Further, NJL model is proposed as alternative phenomenological model of neutrinos.
\end{abstract}

\maketitle

The neutrino is one of the fundamental particle belonging to the lepton. 
It was found in the experiment of $\beta$ decay process.
In present time, it is still mysterious particle \cite{mohap}.
The neutrino has spin 1/2, no electric charge and three flavor ($\nu_e, \nu_\mu, \nu_\tau$). 
It only interacts with a matter in terms of the weak interaction.
Therefore, it is difficult to observe it. 
One important fact of neutrino is that there is only the neutrino which has left handed helicity
(whereas antineutrino has right handed helicity) in nature.
This means that the parity of neutrino is violated.
In present time, we have no exact answer why left handed neutrinos only exist in nature.

In the parity invariant theory, the rate for a process $\alpha \rightarrow \beta (\textrm{with } \alpha \neq \beta)$
is proportional to the transition probability which is symmetric under a reversal of all three-momenta,
$\bm{k} \leftrightarrow - \bm{k}$
\cite{weinberg}. 
However, in the $\beta$ decay process of a polarized cobalt 60 ,
\[ \ ^{60} \textrm{Co} \rightarrow  \ ^{60} \textrm{Ni} + e^- + \bar{\nu}_e, \]
the angular distribution of the electron in the final state were found to be preferentially distribution
\cite{wu}.
Therefore, it has been believed that the weak interaction  violates the parity invariance \cite{lee_yang}.
Note that a fermion moves with the light velocity when the fermion is massless.
This indicates that if neutrino is massless particle, its chirality never change after the Big Ban.

On the other hand, the recent experiment of Super-Kamiokande shows that neutrinos have mass \cite{fukuda} even though
 the mass is very small.
Therefore, it is possible to exist the right handed neutrinos.
Does this fact mean that neutrino is Majorana spinor ?
Due to the seesaw mechanism \cite{seesaw}, the neutrino mass becomes small comparing to the mass scale 
of order $10^{10} \sim 10^{14}$ GeV and more. 
In the seesaw mechanism, super-heavy neutrino is introduced. However, it cannot
be observed since it is indeed heavy!

In this Letter, I propose new scenario of parity violation in neutrino
as well as the origin of neutrino mass.
We consider this problem in different point of view.
This is given in new understanding of the chiral symmetry breaking physics \cite{fhhmt}.
We consider that the neutrino acquires a mass  dynamically in terms of the chiral symmetry breaking.
In this case, the mass may be small since the fermion has essentially massless dispersion relation
\cite{fkht}.

The spontaneous symmetry breaking (SSB) phenomena has been investigated for decades in quantum field theory
\cite{njl,goldstone,klev}.
In particular, one important consequence of the chiral symmetry breaking is that the massless fermion acquires
mass in dynamical way \cite{njl}. 
Recent new investigations \cite{fuj_hir_tak_hepth,nova,fhht,fkht,fhhmt} of the chiral symmetry breaking of
 fermion field theory give us new suggestion for an answer of this problem.

Here, we consider the model of fermion field $\psi$ which is invariant under the chiral transformation,
\begin{equation}
\psi' = e^{i \gamma_5 \theta} \psi.
\end{equation}
In this case, we have the conserved current,
\begin{equation}
j^\mu_5 = - \bar \psi \gamma^\mu \gamma_5 \psi .
\end{equation}
Therefore, the chiral charge, 
\begin{equation}
\hat {\cal Q}_5 = \int \!\!d^3 \bm{x} \,j^0_5(\bm{x},t) ,
\end{equation}
is conserved,
\begin{equation} \left[\hat H, \, \hat {\cal Q}_5 \right] = 0, \label{chiral_ham_rel}\end{equation}
where $\hat H$ is the Hamiltonian.
 Further, let the Lagrangian be invariant under the parity transformation,
\begin{subequations}
\begin{equation}
{\cal P}: \quad t \rightarrow t, \; \bm{x} \rightarrow - \bm{x}
\end{equation}
and
\begin{equation}
 {\cal P}^{-1} \psi(t, \bm{x}) {\cal P} = \gamma_0 \psi(t, -\bm{x}) .
\end{equation}
\end{subequations}
We can easily verify that
\begin{subequations}
\begin{equation}
{\cal P}^{-1} j^\mu_5 (t, \bm{x}) {\cal P} = -j_{5 \mu} (t, -\bm{x})
\end{equation}
and
\begin{equation}
{\cal P}^{-1} \hat {\cal Q}_5 {\cal P} = - \hat {\cal Q}_5 . \label{q5_parity}
\end{equation}
\end{subequations}

In quantum field theory, it is possible that vacuum states are degenerate.
In this case, the spontaneous symmetry breaking should be occurred.
Note, however, that there is no entanglement between degenerate vacuums
 since the vacuums are eigenstate of Hamiltonian.
Further, it is important to note that many vacuums does not exist simultaneously in one physical system.
Therefore, after the symmetry is broken, one of the degenerate vacuums is \textit{only} realized 
as the true vacuum (or physical vacuum) \textit{spontaneously} \cite{fhhmt}. 
Accordingly, after symmetry breaking, the vacuum of the system is \textit{not} trivial vacuum (perturbative vacuum)
 but becomes one broken vacuum in the degenerate vacuums.

Now, we consider the system whose chiral symmetry is spontaneously broken.
 Recent proposal of SSB \cite{fhhmt}
insists that the broken vacuum is not only an eigenstate of the Hamiltonian $\hat H$ but also is that
of the broken charge $\hat {\cal Q}_5$.
Therefore, chiral broken vacuum $|E_\Omega, Q_5 \rangle$ satisfy
\begin{subequations}
\begin{equation}
\hat H |E_\Omega, Q_5 \rangle = E_\Omega |E_\Omega, Q_5 \rangle
\end{equation}
and
\begin{equation}
\hat {\cal Q}_5 |E_\Omega, Q_5 \rangle = Q_5 |E_\Omega, Q_5 \rangle .
\end{equation}
\end{subequations}
From eq.(\ref{q5_parity}), 
\[ {\cal P}^{-1} \hat {\cal Q}_5 {\cal P} |E_\Omega, Q_5 \rangle = - \hat {\cal Q}_5 |E_\Omega, Q_5 \rangle 
   = -Q_5 |E_\Omega, Q_5 \rangle \]
Finally, we have
\begin{equation}
\hat {\cal Q}_5 {\cal P} |E_\Omega, Q_5 \rangle = - Q_5 {\cal P} |E_\Omega, Q_5 \rangle
\end{equation}
This equation shows that one broken vacuum state,
\begin{subequations}
\begin{equation}
|\textrm{vac};+ \rangle = |E_\Omega, Q_5 \rangle,
\label{pbroken_p}
\end{equation}
and its parity transformed state,
\begin{equation}
|\textrm{vac};- \rangle = {\cal P} |E_\Omega, Q_5 \rangle,
\label{pbroken_m}
\end{equation}
\end{subequations}
have different eigenvalue of the chiral charge. 
Note, here again, that the chiral charge is conserved after symmetry breaking because of eq.(\ref{chiral_ham_rel}).

Finally, since one of eq.(\ref{pbroken_p}) and eq.(\ref{pbroken_m}) is only realized after the chiral symmetry breaking,
 the parity transformation property of the physical vacuum of the SSB model is violated
even though the Lagrangian of the model is parity invariant. 

In parity invariant model, the energy dispersion relation of one particle state is symmetric under the reversal,
\begin{equation}
 \bm{k}  \longleftrightarrow - \bm{k},
\label{mom_sym}
\end{equation}
where $\bm{k}$ is momentum of the particle.
It can be understood by the following fact. A left moving particle changes to a right moving particle under the parity
transformation, and vice versa. However, there is no difference between left and right for the case of
 the parity invariant model.
Therefore, both of them have same energy.
For example, the energy dispersion relation of the free fermion is given by
\begin{equation}
E_{\bm{k}} = \sqrt{ \bm{k}^2 + m^2},
\end{equation}
where $m$ is a mass of the particle.
Note that it is symmetric under eq.(\ref{mom_sym}).

 The vacuum of the fermion field model can be constructed in terms
of filling the negative energy particles, which is know as Dirac vacuum.
 Therefore, the vacuum of the parity invariant model is  symmetric under eq.(\ref{mom_sym}).
On the other hand, the momentum  distribution of the vacuum in the parity violation model 
should be \textit{not} symmetric under eq.(\ref{mom_sym}). 

Fortunately, we can find an explicit example for the parity violation of the vacuum as well as breaking
of the chiral symmetry. 
This is massless Thirring model. Although, the Lagrangian of the massless Thirring model is parity and chiral symmetry
invariant model, the true vacuum is not.  The Thirring model is exactly solvable in terms of the Bethe Ansatz Method
\cite{berg_thac}.
For the case of the \textit{massive} Thirring model, the vacuum momentum distribution is symmetric under
 eq.(\ref{mom_sym}) \cite{berg_thac,fkt}.
On the other hand, the exact result of the \textit{massless} Thirring model actually shows that
the chiral symmetry is broken and the chiral broken vacuum of Thirring model is not symmetric
 under eq.(\ref{mom_sym}) whereas the unbroken vacuum is symmetric \cite{fhht}.
 That is, the parity of the physical vacuum in the massless Thirring model is violated.

We, further, can conjecture that the parity of the neutrino is violated
 since one particle state of neutrino should be described in terms of
the broken vacuum. In this respect, we should conclude that neutrinos are only left handed.

In four dimensions, Nambu-Jona-Lasinio (NJL) model is well-known model
 of which the chiral symmetry is spontaneously broken \cite{njl}. Although the NJL model is \textit{not}
 a renormalizable model,  it has been considered as an \textit{effective} model of quarks, and it has been
 believed to describe the lightest pions.
However, the recent investigations show that it should not provide the meson spectrum of quark and antiquark pair
as the Nambu-Goldstone boson \cite{fuj_hir_tak_hepth,nova}. It only provides dynamical mass generation of fermion
as the chiral symmetry breaking in the four dimensions.
In this Letter, we think of it to be the phenomenological model of neutrinos.

The Lagrangian of the NJL model is given by
\begin{equation}
{\cal L} =\dfrac{i}{2} \bar \psi \gamma^\mu \overleftrightarrow{\partial_\mu} \psi
+ G \biggl[ (\bar \psi \psi)^2 + (\bar \psi i \gamma_5 \psi)^2 \biggr] .
\end{equation}
This model is invariant under the parity transformation.
The gap equation of the NJL model in the Bogoliubov method is given by \cite{njl,klev}
\begin{equation}
m^\ast = 4G \int\dfrac{d^3 \bm{k}}{(2 \pi)^3} \dfrac{m^\ast}{E_{\bm{k}}},
\end{equation}
where $E_{\bm{k}} = \sqrt{\bm{k}^2 + m^{\ast 2}}$.
In the four-momentum covariant cutoff scheme \cite{klev}, the gap equation becomes
\begin{equation}
\dfrac{2 \pi^2}{G N_f \Lambda^2} 
 = 1 - \left(\dfrac{m^\ast}{\Lambda} \right)^2 \ln \left[1 + \left(\dfrac{\Lambda}{m^\ast}\right)^2 \right],
\label{njl_gap_eq}
\end{equation}
where $N_f$ is the number of flavor and $\Lambda$ is the cutoff. Note that the dynamical mass should be
small comparing to cutoff $\Lambda$,
 since right hand side of eq.(\ref{njl_gap_eq}) becomes zero in the limit $m^\ast/\Lambda \gg 1$.
Here , it is also important to note that the gap equation of the NJL model depends on the regularization scheme since
the NJL model is unrenormalizable. Therefore, we cannot conclude the smallness of mass in this mechanism.
Nevertheless, it is reasonable to consider that dynamical mass is small.
Here, we should comment on that there is no flavor breaking in mass. It should be discussed in future issue.

When the dynamical mass $m^\ast$ is small comparing to the cutoff $\Lambda$,
the cutoff is given by
\[ \Lambda \sim \sqrt{\dfrac{2 \pi^2}{GN_f}} . \]
Now, we take $G$ as the weak coupling,
\[ G = 1.166 \times 10^{-5} \;\; \textrm{GeV}^{-2}, \]
and $N_f=3$. In this case, the cutoff becomes
\[ \Lambda \sim 751 \;\; \textrm{GeV} .\]
If we take the noncovariant cutoff scheme, the cutoff $\Lambda'$ becomes
\[ \Lambda'=\Lambda/\sqrt{2}=530 \;\; \textrm{GeV}. \]
The cutoff is given by a few times of the Higgs scale $v \sim 246$ (GeV).
Obviously, we need the knowledge of the weak interaction as a local gauge theory for more understanding.
However, we can expect that neutrinos should be  examined phenomenologically in terms of the NJL model.

It should comment on the Nambu-Goldstone (NG) boson \cite{njl,goldstone}.
It may be still believed that the NG boson always appear after SSB.
However, recent realization of the SSB show that the NG boson does \textit{not} always appear after
SSB in fermion field theory \cite{fuj_hir_tak_hepth,nova,fhht,fkht,fhhmt}.
 The important point is that Goldstone theorem does not predict any scalar boson.
It only states that there is momentum zero mode \textit{if the scalar field exists}.
However, the existence of the scalar field is an strong physical assumption in fermion field theory
like the NJL and Thirring models. According to the analysis of \cite{nova,fhht,fkht,fhhmt,fuj_hir_tak_hepth},
there should be \textit{no} NG boson even after SSB have been occurred in NJL and Thirring models.
Further, there should be no bound states of fermions since the coupling constant is small.
In this respect, we need not to introduce any alternative new particle in this scenario.

I gratefully thank to Prof. Fujita for discussions and useful comments.

\end{document}